# Bitemporal Dynamic Sinai Divergence: An Energetic Analog to Boltzmann's Entropy?

## Otto E. Rossler[1] and Ramis Movassagh[2]

[1]Division of Theoretical Chemistry, University of Tübingen, Auf der Morgenstelle 8, 72076 Tübingen, F.R.G. and International Institute for Advanced Studies, IIAS University of Windsor, Ont., Canada

[2]ETH and Collegium Helveticum, Zurich, Switzerland

Abstract

Sinai chaos is characterized by exponential divergence between neighboring trajectories of a point billiard. If the repulsive potential of the finite-diameter fixed particle in the middle of the table is made smooth, the Sinai divergence persists with finite measure. So it does if the smooth potential is made attractive. So it still does if the potential is in addition made time-dependent (periodic). Then a systematic decrease in energy of the moving particle can be predicted to occur in both time directions for a long time. If so, classical entropy acquires an analog in real space.

A "Sinai gas" consists of a single point billiard moving on a quadratic table with a fixed repelling central disk – while opposing pairs of sides of the table are identified ("flat torus"). Sinai proved that neighboring parallel paths exhibit exponential divergence for almost all initial conditions[1]. The preconditions for chaos of Li and Yorke [2] are therefore fulfilled. Remarkably, the Sinai divergence of spatial trajectories occurs in both directions of time.

The result remains valid if the repulsive potential of the central disk is made smooth rather than discontinuous, although the chaotic solutions then no longer are the only ones with full measure, cf. [3]. The Sinai divergence remains valid even if the central disk is made smoothly attracting rather than repelling (with an 1/r potential, say). Hereby the initial speed has to be made large enough to rule out circular and elliptic motions about the central disk or point.

Finally, Sinai divergence persists even if the attractive potential of the central disk or point is made time-dependent (oscillatory). The divergence then occurs no longer only in space (static Sinai divergence) but also in time (dynamic Sinai divergence), and so in both directions of time.

As long as the amplitude of the assumed slow oscillation is very weak, it goes without saying that the trajectories remain close to the static case – so that the dynamic Sinai divergence indeed exists in both directions of time. To better understand what happens when the temporal modulation is marked, Sinai's "gluing method," as it may be called, can be applied to a representative subset of situations. Specifically, a sequence of equal-angle, equal-distance, equal-momentary-potential encounters between the moving point and the breathing radial-symmetric attracting funnel with its flat bottom is assumed without loss of generality. The only difference between the otherwise equal pairwise members is that their potential's change in time is positive in the one twin case and negative in the other. Although the attractive potential has the same strength as the funnel is entered by the moving particle (whereby a careful determination of the correct finite entering level is presupposed since the "flat outer part" of the funnel extends to infinity), it does not remain constant but either increases or decreases in strength compared to the other, during the subsequent further approach toward the center and the subsequent veering-away from it. In other words the fall is steeper than the climb-up. Therefore the two equiprobable subcases that make in the assumed simplified sequence have unequal effects: the "steepening" ones have a stronger cumulative effect over time than the "flattening" ones.

That is, the steepening effects take away more energy from the moving particle than the flattening ones add. Hence the particle predictably suffers a net decrease of its momentum in proportion to the travel time. Eventually, the trajectory must cease being approximately straight and enter a low-temperature "pseudo-equilibrium" that is virtually impossible to leave again in finite (forward) time.

The prediction arrived at of a systematic change in energy with time of the moving point particle is rather surprising, for it applies in either direction of time for a freshly picked initial condition. This behavior strongly resembles that of entropy in statistical mechanics (cf., for example, [4]). Simulations will be necessary to confirm or refute the prediction. Since the phenomenon, if it is real, was possibly first encountered in an empirical context by Vesto M. Slipher in 1914 [5], it deserves to be called slentropy (or "Slipher entropy").

## Acknowledgments

Paper dedicated to Mohamed ElNaschie on the occasion of his 60[th] birthday. O.E.R. in addition thanks Frank Kuske for a discussion, Kam-To Leung, Ji-Huan He, Dietrich Hoffmann and Peter Weibel for stimulation and Florian Grond for a preliminary simulation of a more complex system. For J.O.R.